\documentclass[12pt,preprint]{aastex}

\slugcomment{\underline{Accepted for}: ApJL \hfill \underline{Version}: \today}

\shorttitle{Pulsating Reverse Detonation models}
\shortauthors{Bravo\& Garc\'\i a-Senz}

\usepackage{graphicx}
\usepackage{color}

\begin{document}

\title{Beyond the bubble catastrophe of Type Ia supernovae: Pulsating Reverse 
Detonation models}
\author{Eduardo Bravo\altaffilmark{1,2}, 
Domingo Garc\'\i a-Senz\altaffilmark{1,2}}
\altaffiltext{1}{Dept. F\'\i sica i Enginyeria Nuclear, Univ. Polit\`ecnica de
Catalunya, Diagonal 647, 08028 Barcelona, Spain;   
eduardo.bravo@upc.edu domingo.garcia@upc.edu}
\altaffiltext{2}{Institut d'Estudis Espacials de Catalunya, Barcelona, Spain}

\begin{abstract}
We describe a mechanism by which a failed deflagration of a Chandrasekhar-mass 
carbon-oxygen white dwarf can turn into a successful 
thermonuclear supernova explosion, without invoking an {\sl ad hoc} high-density
deflagration-detonation transition. Following a pulsating phase, an accretion 
shock develops above a core of $\sim1$~M$_\sun$ composed of carbon and oxygen, 
inducing a converging detonation. A three-dimensional simulation of the
explosion produced a kinetic energy of $1.05\times10^{51}$~ergs and
$0.70$~M$_\sun$ of
$^{56}$Ni, ejecting scarcely $0.01$~M$_\sun$ of C-O moving at low velocities.
The mechanism works under quite general conditions and 
is flexible enough to account for the diversity of normal Type Ia supernovae. 
In given conditions the detonation might not occur, which would 
reflect in peculiar signatures in the gamma and UV-wavelengths.
\end{abstract}


\section{Introduction}

In order to measure cosmic distances with the precision required to determine
the equation of state of 
the dark energy component of our Universe, it is necessary to understand the 
physics of Type Ia supernovae (SNIa). From the theoretical point of view, the 
accepted model of SNIa consists of a Chandrasekhar-mass white dwarf (WD) 
that accrets matter from a close 
binary companion. This scenario accounts for the SNIa sample homogeneity, the 
lack of hydrogen in their spectra, and its detection in elliptical galaxies. 
There 
are two main ingredients of the standard model that are still poorly known: 
the precise configuration and evolution of the binary system prior 
to thermal runaway of the WD, and the explosion mechanism. 
In spite of 40 years of theoretical efforts dedicated to understand the 
mechanism behind SNIa, realistic simulations are still unable to provide a 
satisfactory description of the thermonuclear explosion. Nowadays, there is 
consensus that the initial phases of the explosion involve a subsonic 
thermonuclear flame (deflagration), whose propagation competes with the 
expansion of the WD.  After a while the corrugation of the flame front  
induced by hydrodynamic instabilities leads to an acceleration of the effective 
combustion. The nature of the events that follow is currently 
under debate between advocates of a transition to a supersonic detonation front 
and those defending that the flame remains subsonic. 
Recent 
three-dimensional (3D) models calculated by different groups have shown that both 
explosion mechanisms display positive as well as serious weak points. 
Pure deflagrations always give final kinetic energies that fall short 
of $10^{51}$~ergs and leave too much unburnt carbon and oxygen (C-O) close to the 
center \citep{g03,hn,r02}. Both results are at odds with observational 
constraints. 
Leaving aside pure deflagrations, \citet{g03} proposed that it would be
necessary 
to assume that the turbulent flame triggers a detonation. 
On the other hand, the deflagration-detonation transition (DDT) had to be
postulated {\it ad hoc} \citep{a94,h95}, because current numerical experiments
disfavour such a transition in exploding WDs \citep{n99}. 

In \citet{bg05} it was sketched a mechanism through which a delayed detonation
might naturally arise: the Pulsating Reverse Detonation (PRD) model. In 
the PRD paradigm the explosion proceeds in three steps: 1) an initial 
pre-conditioning phase whose result is the inversion of the chemical structure 
of the progenitor WD, 2) the formation of an accretion shock that confines the 
fuel and, 3) the launch of an inward moving detonation. In this Letter we
present the results of 3D simulations of the PRD model. In the next section we
describe our simulations and analyse the properties of the
accretion shock, while in the final section we speculate about the implications of
this new paradigm of SNIa. A more detailed report with
additional calculations of the PRD model will be published elsewhere
(Bravo \& Garc\'\i a-Senz, in preparation).

\section{The Pulsating Reverse Detonation model}

Our initial model consists of a $1.38$~M$_\sun$ C-O WD. 
The hydrodynamic evolution started with the ignition of 6 
sparks [model B06U in \citet{gar05}] incinerating 
M$_\mathrm{def}=0.18$~M$_\sun$ in one 
second, while releasing $2.5\times10^{50}$~ergs of nuclear energy which led to the 
pulsation of the star. 

\begin{figure}
\plotone{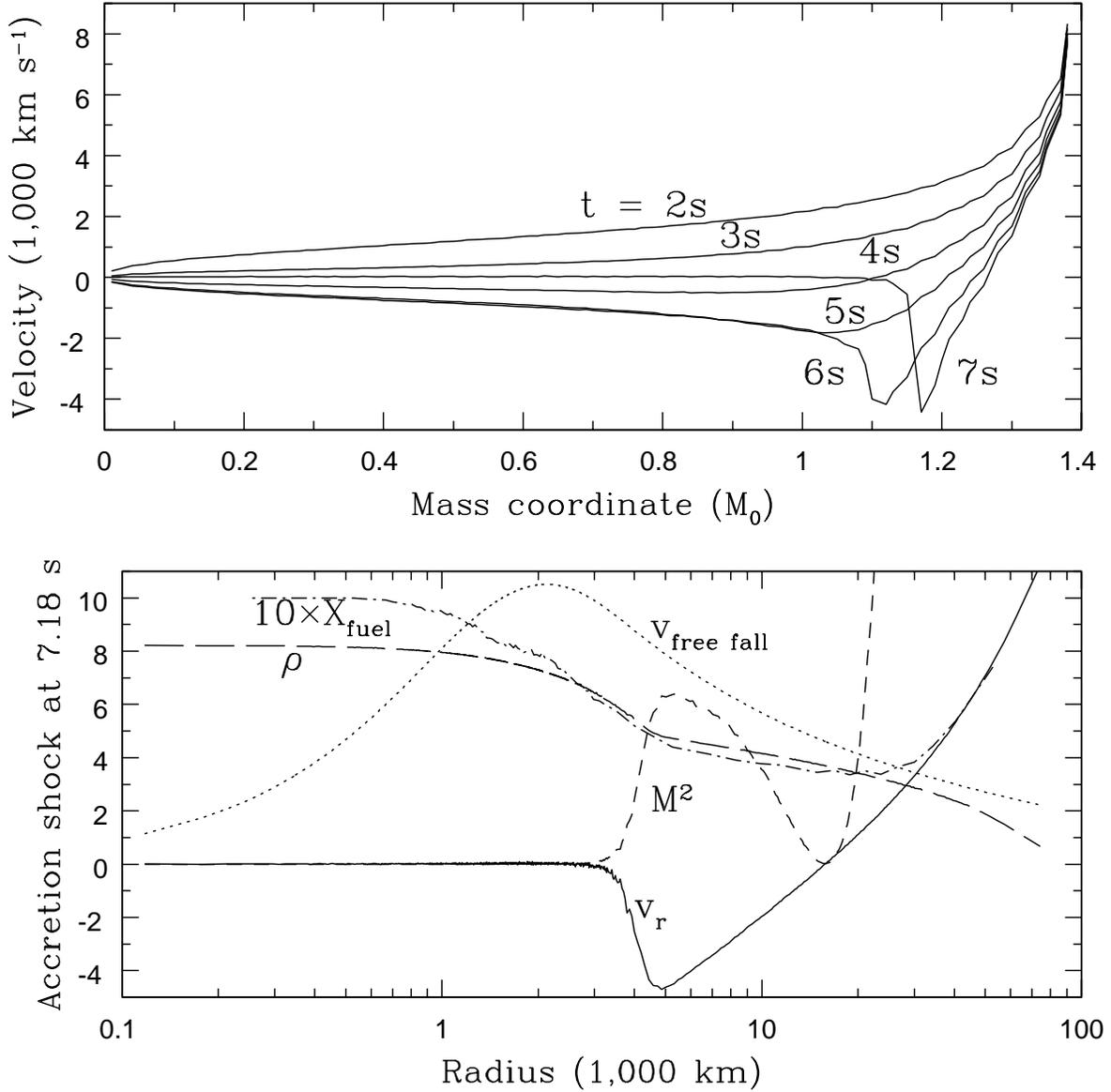}
\caption{ 
Formation of the hydrostatic core and accretion shock, whose structure is shown 
in the bottom panel. In the top panel there is depicted the evolution of the 
angle-averaged velocity profile during the pulsation of the WD, 
computed with a 3D hydrocode. The bottom panel shows the 
angle-averaged structure at the end of the pulsation, 7.18 s after 
thermal runaway, as a function of radius. It can be seen the hydrostatic core 
up to $\sim2,500$~km, the accretion 
shock at $\sim5,000$~km, and the expanding atmosphere above $\sim20,000$~km. 
The curves represent  
the angle-averaged radial velocity, $v_\mathrm{r}$ in units of 
$1,000$~km$\cdot$s$^{-1}$, 
free fall velocity, $v_\mathrm{free~fall}$ in the same units, square of the 
Mach number, $M^2$,
logarithm of density, $\rho$ in g$\cdot$cm$^{-3}$, and mass fraction of C-O, 
$X_\mathrm{fuel}$
}\label{fig1}
\end{figure}

The details of the first phase, that spans the first two seconds after thermal 
runaway, have been known for some time \citep{ple04,l05}. 
Even though the precise configuration at 
thermal runaway is difficult to determine, current works suggest a multipoint ignition in which the first 
sparks are located slightly off-center \citep{gsw95,w04}. If the number of 
sparks is too 
small the nuclear energy released is not enough to unbind the star and the 
explosion fails. This is known as the "single-bubble catastrophe" \citep{l05}. 
These bubbles float to the surface before the combustion wave can propagate 
substantially \citep{ple04,gar05} and the star remains energetically bound. 
This behaviour 
produces a composition inversion, i.e. the fuel, composed of cold C-O, 
fills the internal volume while the ashes of the initial combustion, 
mostly hot iron and nickel, are scattered around (Fig. \ref{fig1}). In the 
3D calculations the energy resides for the most part in the 
outer $0.15$~M$_\sun$. Hence, the expanding motion of the external material is 
decoupled from the rest of the structure, and a pulsation starts. 

The second phase of the explosion begins when the deflagration quenches due to 
expansion and ends when an accretion shock is formed by the impact of the 
infalling material. At the end 
of this phase, the inner $1.0$~M$_\sun$ C-O rich core adopts 
an equilibrium configuration inertially confined by an 
accretion shock (Fig. \ref{fig1}).
Before describing the events that ensue to the 
formation of the accretion shock, we will perform an analysis of its 
properties, based on the structure of the shock resulting from our simulation 
of the first pulsation of the WD. We show that a detonation is expected to start
a few thousand kms below the accretion shock and that, once
formed, it cannot be quenched due to a rapid expansion of the core, which 
remains confined due to the large impact pressure of the accreting matter.
This analysis is intended to provide the physical basis for the formation of the
detonation in order to support the results of our hydrodynamical calculations,
which might be affected by numerical resolution.

The structure of the WD at t = 7.18 s can be seen in Fig. \ref{fig1}.
Below $\sim15,000$~km matter is flowing in 
towards the hydrostatic core. The accretion shock starts at
$r_\mathrm{sh}\sim~5,000$~km 
(lagrangian mass $\sim1.17$~M$_\sun$), where the inwards velocity of the 
infalling matter is $v_\mathrm{r}\sim5,000$~km$\cdot$s$^{-1}$.
The density just above the 
accretion shock is $\rho_0=8.3\times10^4$g$\cdot$cm$^{-3}$, while that 
of shocked matter
rises to $\rho_\mathrm{sh}=4\rho_0=3.3\times10^5$~g$\cdot$cm$^{-3}$. Once 
formed, the accretion shock remains 
confined close to the hydrostatic core due to the large impact pressure of the 
infalling matter compared to the gas pressure behind the shock: $\rho v^2/p = 
\gamma M^2\sim12$, where $\rho v^2$ is 
the impact pressure, $p$ is the gas pressure, $\gamma=5/3$ is the adiabatic 
coefficient, and $M$ is the Mach number. We can reformulate this point in a more 
quantitative way as follows. The rate of mechanical energy deposition at the 
accretion shock is given by $\dot{\varepsilon}_\mathrm{mec} = 2\pi r^2\rho_0
v^3$, which can be compared to the rate 
of nuclear energy released in a combustion front propagating at velocity $c$, 
$\dot{\varepsilon}_\mathrm{nuc} = 4\pi r^2 \rho_\mathrm{sh} cq$, where 
$q\sim5.8\times10^{17}$~erg$\cdot$g$^{-1}$ is the difference in nuclear 
binding energy between matter composed of C-O and $^{28}$Si. Equating both 
energy rates one obtains $c=270$~km$\cdot$s$^{-1}$. This 
is roughly the flame velocity that would be required to revert the bulk inward 
motion of the accreting matter. It turns out that this velocity is 
$\sim0.1 v_\mathrm{sound}$, that is much 
larger than the maximum velocity of a stable deflagration at
$\rho_\mathrm{sh}$ \citep{k88}, 
$v_\mathrm{sound}$ being the local sound velocity. Furthermore, due to the converging 
nature of the combustion wave the fuel has nowhere to expand, ensuring that 
the flame will not be quenched. 
 
There is an additional condition for detonation initiation: the 
nuclear time scale must be lower than the hydrodynamical timescale. The 
temperature attained at the accretion shock can be estimated from the 
Rankine-Hugoniot conditions for a strong shock and the ideal gas equation of 
state, giving 
$T_\mathrm{sh} = \left(3/16\right)\left(\mu/\mathrm{k_B N_A}\right)v^2$,
where $\mu$ is the mean molar mass, $\mathrm{N_A}$ is Avogadro's number, and
$\mathrm{k_B}$ is the Boltzmann 
constant. In matter composed of completely ionized C-O, 
$\mu=1.75$~g$\cdot$mol$^{-1}$, which gives $T_\mathrm{sh}=10^9$~K. At the 
density of the shock, this temperature 
is not high enough to burn in less than a hydrodynamical 
time, $\tau_\mathrm{hyd}=446/\sqrt{\rho_\mathrm{sh}} = 0.78$~s, thus the 
shocked matter must be compressed 
further along its path towards the surface of the hydrostatic core before 
detonating. Due to the high accretion rate, the shocked gas remains optically 
thick and photons are trapped in the infalling matter, implying that 
radiative cooling is inefficient and the flow is radiation dominated, i.e. the
adiabatic coefficient goes down to $\gamma=4/3$. Assuming an adiabatic 
evolution of 
this radiation-dominated shocked matter, $T\propto\rho^{1/3}$, one can compare 
the local values of the nuclear 
timescale, $\tau_\mathrm{nuc}$, and $\tau_\mathrm{hyd}$ in order to find the 
radius at which explosive ignition is reliable. With the additional
approximations of steady state and spherical symmetry
the structure of the shocked flow between the 
accretion shock and the core can be obtained by solving the following 
set of equations: 
$e = v^2/2 + p/\left[\rho\left(\gamma-1\right)\right] - GM/r =
\mathrm{constant}$ (conservation of energy),
$\rho v r^2 = \mathrm{constant}$ (conservation of mass), and 
$p\propto\rho^{4/3}$ (adiabatic evolution), starting from the 
physical state behind the shock, $\rho_\mathrm{sh}$, $T_\mathrm{sh}$, 
and $v_\mathrm{sh}=v/4=1,250$km$\cdot$s$^{-1}$. We have found that 
$\tau_\mathrm{nuc} < \tau_\mathrm{hyd}$ 
at $r_\mathrm{det} =3,400$~km and $\rho_\mathrm{det} =
1.3\times10^6$~g$\cdot$cm$^{-3}$, i.e. a detonation is able to 
form $0.12$~M$_\sun$ inside the accretion shock. It turns out that the ignition 
of a detonation critically
depends on the amount of mass burnt during the deflagration phase: the larger
M$_\mathrm{def}$ the more difficult is the formation of a detonation.
In another model that burned M$_\mathrm{def}=0.29$~M$_\sun$ subsonically, the 
temperature and density behind the shock rised only to $4.5\times10^8$~K 
and $1.4\times10^5$~g$\cdot$cm$^{-3}$. Eventually, one can expect that 
increasing slightly the value of M$_\mathrm{def}$ the 
physical conditions behind the shock would not allow the formation of a stable 
burning front and the explosion would finally fail. We expect such a failure 
for a narrow range of M$_\mathrm{def}\sim 0.29-0.35$~M$_\sun$, in which
case the event would not resemble a SNIa.

\begin{figure}
\plotone{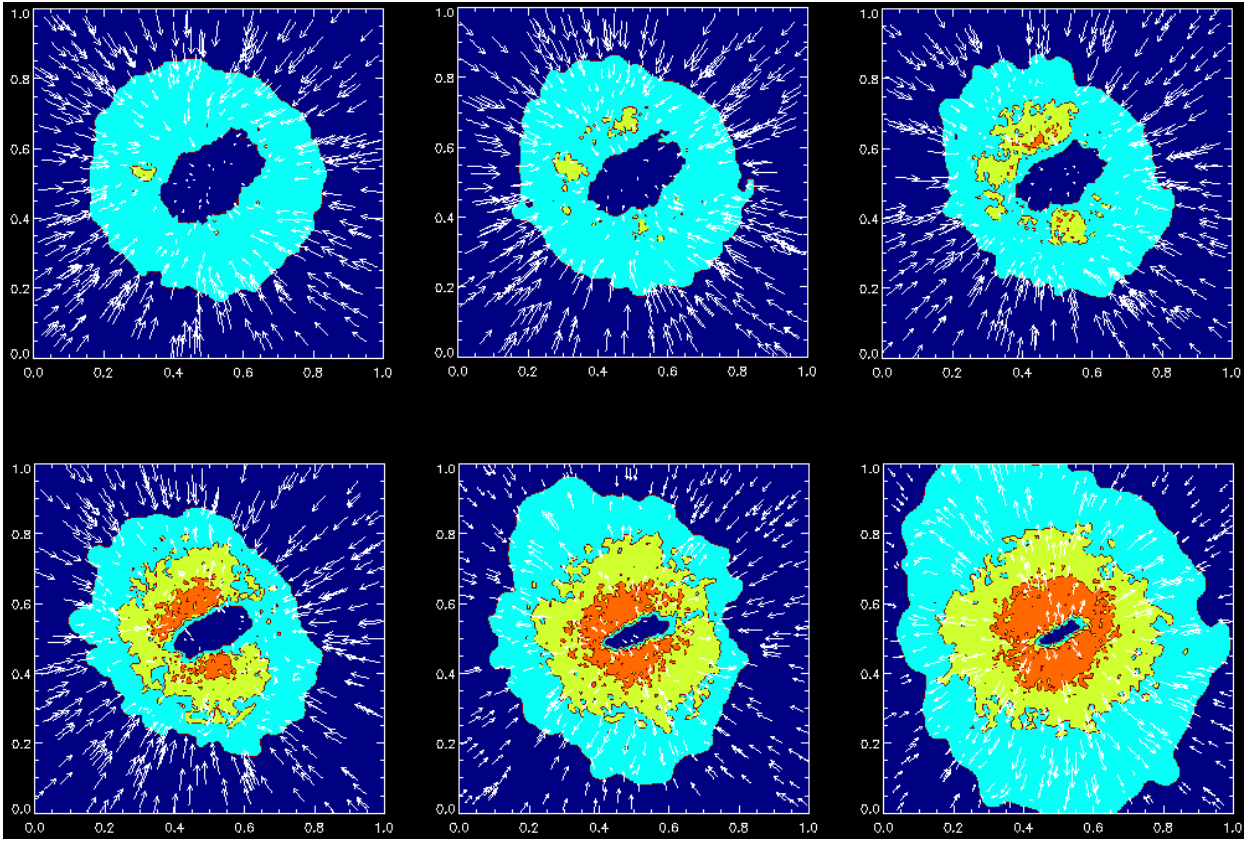}
\caption{
Evolution of the velocity field and isotemperature contours during the
detonation phase in a slice of side 10,000 km, which 
encloses $\sim85$\% of the mass of the WD. The snapshots are shown in time 
steps of 0.1 s since the formation of the detonation. The temperature contours 
begin with $T=5\times10^8$~K and
continue in steps of $2\times10^9$~K. 
The detonation starts in several isolated hot spots (first two snapshots), 
afterwards propagates rapidly in azimuthal direction
(third and fourth snapthots), and finally inwards in radial
direction (last two snapshots). In the last snapshot it is apparent the vigorous
expansion of the detonated material, which sends 
an inwards moving rarefaction wave 
that weakens the detonation front and finally quenches burning. 
The maximum velocities shown at each time 
are, from left to right and top to bottom: 6,000, 5,808, 6,844, 7,202, 10,096, 
and 10,986 km~s$^{-1}$. The values of the maximum resolution at each snapshot 
are: 20, 18, 15, 12, 11, and 11 km .[See the electronic edition of the Journal
for a color version of this figure.]
}\label{fig2}
\end{figure}

\begin{figure}
\plotone{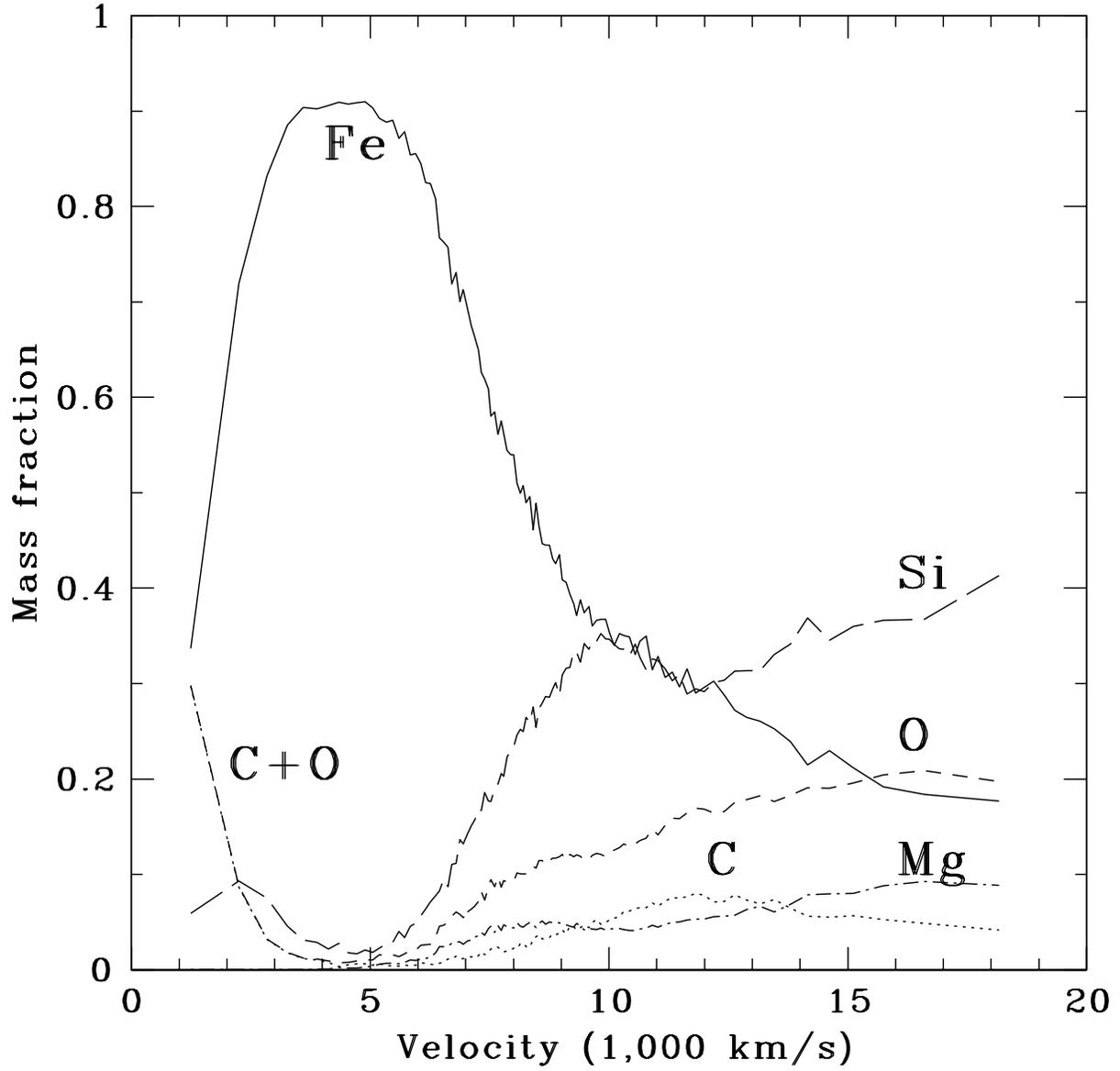}
\caption{ 
Final distribution of the main chemical elements (after
radioactive decays) as a function of final velocity
}\label{fig3}
\end{figure}

The third phase of the explosion starts when the converging reverse detonation 
wave is launched. This phase has been computed with our 3D 
hydrodynamical code, the same as in 
\citet{g99} although with higher resolution (here, we used 250,000 particles). 
In our model the detonation starts naturally as a result of the compressional
heating caused by the accretion shock and, 
once initiated, the detonation is self-sustained by the burning of fuel either
to intermediate-mass or to Fe-group elements.
During the 
first 0.3~s after detonation ignition the 
accretion shock remains stationary, close to the core (Fig.\ref{fig2}). 
Afterwards, the 
overpressure generated by the nuclear energy released pushes outwards the 
accretion 
shock, that detaches from the core, and the detonated matter starts to expand
with large velocities. The expansion weakens the detonation and finally the
burning quenches. As a result, the detonation burns 
all the fuel except for a tiny region at the center. 
In the outer layers of the hydrostatic core the density is low enough to allow 
incomplete burning and leave a composition rich in intermediate mass elements 
(Fig.\ref{fig3}) while in the inner regions the burning 
proceeds all the way up to $^{56}$Ni (the central density at the moment of 
formation of the detonation is $10^8$~g$\cdot$cm$^{-3}$). The simulation was
ended when the elapsed time was 5176~s after initial thermal runaway.

How do the hydrodynamical simulations of the PRD model compare to observations 
of SNIa? Nowadays, a detailed comparison with existing data 
is not possible because multidimensional spectral and photometric
codes are not fully developed. However, in general terms, our results match 
quite satisfactorily basic SNIa 
observational constraints. The mechanical structure of the ejecta retains a 
high degree of spherical symmetry and is chemically stratified, although there
is some small-scale mixing and clumping of chemical elements. The final kinetic 
energy is $1.05\times10^{51}$~ergs, and the total amount of $^{56}$Ni produced 
in the event is $0.70$~M$_\sun$, in 
good agreement with what is demanded by observations of typical SNIa
\citep{bk}. 
There remain only $0.06$~M$_\sun$ of unburned C, most of it moving at high 
velocities. 
This carbon, as well as the remaining unburned O (Table\ref{tab1}) would be 
hardly detectable around the epoch of maximum light \citep{blh03}. 
The total amount of C-O moving at low velocities is $0.01$~M$_\sun$, that is
within the limits derived by \citet{koz} from late-time spectra.
Intermediate mass elements moving at
velocities $>8,000$~km$\cdot$s$^{-1}$ are abundant at the photospheric radius 
at maximum light, as required by observations. 

\section{Discussion}

In the beginning, the motivation of our calculation was to reproduce a 3D
analogue
of the Pulsating Delayed Detonation (PDD) model introduced by \citet{i74} and
developed by \citet{k91}.
In both the PDD and the PRD models there is an initial
unsuccessful deflagration phase leading to a pulsation of the WD. However, in the
PDD model the incinerated matter stays at the center, and there is neither 
composition inversion nor accretion shock. Thus, the mechanism of 
formation of a detonation and the final result is quite different in both kind
of models.

The PRD paradigm is capable of producing a variety of outcomes, which 
could account for a part of the observed diversity of SNIa. This variety derives
from differences in the mass of the hydrostatic core at the moment of formation
of the accretion shock, and is allowed by the general validity of the mechanism
of initiation of the detonation. We have performed a preliminary exploration of
the sensitivity of the explosion properties to the mass of the hydrostatic core 
by means of 1D hydrodynamic calculations
(Bravo \& Garc\'\i a-Senz, in preparation). For reasonable choices of the core
mass, the final kinetic 
energies range from $0.72\times10^{51}$ up 
to $1.21\times10^{51}$~erg, and the $^{56}$Ni masses from 0.35 to 
$0.88$~M$_\sun$. This range  
would translate in differences of up to 1~mag, 
in fairly agreement with the observational range of absolute magnitudes of 
{\sl normal} SNIa (from $M_\mathrm{B} = -18.62$ for SN1983G to $M_\mathrm{B} = 
-19.69$ for SN1997bp). 
 
The origin of the variation of core masses could be due to a randomness of the 
number of sparks igniting initially at the centre of accreting WDs. A large 
number of sparks would imply a larger nuclear energy release during the 
deflagration phase, thus a smaller fuel-rich hydrostatic core after initial 
pulsation and a more loose structure at the moment of accretion shock 
formation, leading to lower values of the final kinetic energy and $^{56}$Ni mass. 
The initial number of sparks could also be related to the rotation of the WD
\citep{k05}, a lower number being favoured in slow rotators, 
opening an interesting connection between the pre-supernova evolution of the 
binary system and the explosion properties.

We have identified two possible observational tests of the PRD paradigm, both of
them related to the eventual failure to develop a detonation (thus not giving a
SNIa-like phenomenon) if the mass burned  
prior to pulsation lies in the range $\sim 0.29-0.35$~M$_\sun$. 
First, even in the absence of a detonation 
the outermost $\sim 0.23-0.28$~M$_\sun$ (half of which 
$^{56}$Ni) would have 
enough energy to escape from the WD. Due to the low column density, the 
radioactive photons emitted in the disintegration of $^{56}$Ni and 
$^{56}$Co would not be 
efficiently thermalized to optical wavelengths. We have estimated that during the first 30-40~days
the gamma-ray spectrum would be similar to the one predicted for 
sub-Chandrasekhar models \citep{m04}, but after $\sim50$~days, when the ejected
matter becomes optically thin, the continuum and 511~keV lines would 
turn much
fainter, due to the smaller mass of $^{56}$Ni synthesized. Second, 
the small amount of mass ejected
would not be enough to destroy the binary system. The stellar 
remnant left back by the failed supernova would be characterized for some time 
by a high 
luminosity in the UV, an exotic surface chemical composition (rich in C, O, Fe
and Ni), an eccentric orbit, and a mass $\sim1.10-1.15$~M$_\sun$. The 
detection of such objects and the measurement of its properties might provide
important information about the final fate of WDs experiencing ignition in a few
bubbles.

\acknowledgments

This work has been supported by DURSI of the Generalitat de Catalunya and 
Spanish DGICYT grants AYA2000-1785, AYA2001-2360 and AYA2002-04094. In loving 
memory of Mar\'\i a del Llano and L. Alexandra.

\clearpage
\centering
\begin{deluxetable}{ccccccc}
\tabletypesize{\scriptsize}
\tablecaption{\label{tab1}Results of the 3D simulation of the PRD model}
\tablecolumns{7}
\tablewidth{0pt}
\tablehead{
\colhead{$K$} &
\colhead{M $(\mathrm{C})$} &
\colhead{M $(\mathrm{O})$} &
\colhead{M $(\mathrm{Mg})$} &
\colhead{M $(\mathrm{Si})$} &
\colhead{M $(^{56}\mathrm{Ni})$} &
\colhead{$Y_\mathrm{e}$} \\
\colhead{$(10^{51}~\mathrm{erg})$} &
\colhead{(M$_\sun$)} &
\colhead{(M$_\sun$)} &
\colhead{(M$_\sun$)} &
\colhead{(M$_\sun$)} &
\colhead{(M$_\sun$)} &
\colhead{$(\mathrm{mol}\cdot\mathrm{g}^{-1})$} 
}
\startdata
1.05 & 0.06 & 0.15 & 0.05 & 0.28 & 0.70 & 0.4976 \tablenotemark{a} \\
\enddata
\tablenotetext{a}{Mean final electron mole number}
\end{deluxetable}

\clearpage

\clearpage

\clearpage


\begin{thebibliography}{}
\bibitem[Arnett and Livne(1994)]{a94} Arnett, W.D., \& Livne, E. 1994, \apj,
427, 330
\bibitem[Baron, Lentz, and Hauschildt(2003)]{blh03} Baron, E., Lentz, E.J., \& 
Hauschildt, P.H. 2003, \apj, 588, L29
\bibitem[Branch and Khokhlov(1995)]{bk} Branch, D., \& Khokhlov, A. 1995, 
\physrep, 256, 53
\bibitem[Bravo and Garc\'\i a-Senz(2005)]{bg05} Bravo, E., \& Garc\'\i a-Senz, 
D. 2005, in Cosmic Explosions. On the 10th Anniversary of SN1993J (IAU Coll.
192), eds. J. Marcaide \& K.W. Weiler (Berlin: Springer), 339-344 
(astro-ph/0401230)
\bibitem[Gamezo et al.(2003)]{g03} Gamezo, V.N., Khokhlov, A.M., Oran, E.S., 
Ctchelkanova, A.Y., \& Rosenberg, R.O. 2003, Science, 299, 77
\bibitem[Garcia-Senz and Bravo(2005)]{gar05} Garc\'\i a-Senz, D., Bravo, E., 
2005, \aap, 430, 585 
\bibitem[Garc\'\i a-Senz, Bravo and Woosley(1999)]{g99} Garc\'\i a-Senz D.,
Bravo, E., \& Woosley, S.E. 1999, \aap, 349, 177
\bibitem[Garc\'\i a-Senz and Woosley(1995)]{gsw95} Garc\'\i a-Senz D., \& 
Woosley, S.E. 1995, \apj, 454, 895
\bibitem[Hillebrandt and Niemeyer(2000)]{hn} Hillebrandt, W., \& Niemeyer, J. 
2000, \araa, 38, 191 
\bibitem[H\"oflich, Khokhlov, and Wheeler(1995)]{h95} H\"oflich, P., Khokhlov,
A., \& Wheeler, J.C. 1995, \apj, 444, 831
\bibitem[Ivanova, Imshennik, and Chechetkin(1974)]{i74} Ivanova, L.N.,
Imshennik, V.S., \& Chechetkin, V.M. 1974, \apss, 31, 497
\bibitem[Khokhlov(1988)]{k88} Khokhlov, A. 1988, \apss, 149, 91
\bibitem[Khokhlov(1989)]{k89} Khokhlov, A. 1989, \mnras, 239, 785
\bibitem[Khokhlov(1991)]{k91} Khokhlov, A. 1991, \aap, 245, L25
\bibitem[Kozma et al.(2005)]{koz} Kozma, C., Fransson, C., Hillebrandt, W., 
Travaglio, C., Sollerman, J., Reinecke, M., Röpke, F.K., Spyromilio, J. 2005,
\aap, 437, 983
\bibitem[Kuhlen, Woosley, and Glatzmaier(2005)]{k05} Kuhlen, M., Woosley, S.E., 
Wunsch, S. 2005, submitted to \apj (astro-ph/0509367) 
\bibitem[Livne, Asida, and H\"oflich(2005)]{l05} Livne, E., Asida, S.M., \&
H\"oflich, P. 2005, \apj, 632, 443
\bibitem[Milne et al.(2004)]{m04} Milne, P.A., et al. 2004, \apj, 613, 1101
\bibitem[Niemeyer(1999)]{n99} Niemeyer, J.C. 1999, \apj, 523, L57
\bibitem[Plewa, Calder, and Lamb(2004)]{ple04} Plewa, T., Calder, A. C., Lamb, 
D. Q., 2004, \apj, 612, L37
\bibitem[Reinecke, Hillebrandt, and Niemeyer(2002)]{r02} Reinecke, M.,
Hillebrandt, W., \& Niemeyer, J. 2002, \aap, 391, 1167
\bibitem[Woosley, Wunsch, and Kuhlen(2004)]{w04} Woosley, S.E., Wunsch, S., \& 
Kuhlen, M. 2004, \apj, 607, 921 
\end{thebibliography}
\end{document}